%% file: main.tex
\documentclass[manuscript, screen, noacm]{acmart}

\usepackage{enumitem}
\usepackage{subcaption}

\AtBeginDocument{%
  }

\acmISBN{978-X-XXXX-XXXX-X/XX/XX}

\renewcommand\footnotetextcopyrightpermission[1]{}
\begin{document}
\emergencystretch=2em
\sloppy
\hbadness=10000
\vbadness=10000
\hfuzz=50pt
\vfuzz=50pt

\title{\name{}: Computer Architecture Simulation as a Service}
\subtitle{\normalsize{}}



\newcommand{\takeaway}[1]{
\noindent\rule{\linewidth}{0.1pt}
\par\nobreak\noindent\textbf{Takeaways:}
#1
\vspace{-2mm}
\par\nobreak\noindent
\rule{\linewidth}{0.1pt}
}

\newcommand{\name}{ArchSim}
\newcommand{\yifan}[1]{\textcolor{blue}{[Yifan: #1]}}
\newcommand{\wenhan}[1]{\textcolor{orange}{[Wenhan: #1]}}
\newcommand{\sabila}[1]{\textcolor{purple}{#1}}
\newcommand{\kate}[1]{\textcolor{teal}{[Kate: #1]}}

\author{Sabila Al Jannat}
\email{sajannat@wm.edu}
\affiliation{%
  \institution{William \& Mary}
  \city{}
  \state{}
  \country{USA}}
\author{Wenhan Lyu}
\email{wlyu@wm.edu}
\affiliation{%
  \institution{William \& Mary}
  \city{}
  \state{}
  \country{USA}}
\author{Le Khanh Trinh Mai}
\email{u1492442@utah.edu}
\affiliation{%
  \institution{The University of Utah}
  \city{}
  \country{USA}}
\author{Huizhi Zhao}
\email{hzhao1001@gmail.com}
\affiliation{%
  \institution{William \& Mary}
  \city{}
  \country{USA}}
\author{Zhuoyan Zheng}
\email{zzheng11@wm.edu}
\affiliation{%
  \institution{William \& Mary}
  \city{}
  \country{USA}}
\author{Katherine E. Isaacs}
\email{kisaacs@sci.utah.edu}
\affiliation{%
  \institution{The University of Utah}
  \city{}
  \country{USA}}
\author{Yifan Sun}
\email{ysun25@wm.edu}
\affiliation{%
  \institution{William \& Mary}
  \city{}
  \country{USA}}




\input{sections/00_Abstract}



\acmConference{}{}{}{}
\maketitle

\input{sections/01_Introduction}
\input{sections/02_Motivation}
\input{sections/03_System_Overview}
\input{sections/04_Evaluation}
\input{sections/05_Case_studies}

\input{sections/07_Related_work}
\input{sections/08_Conclusion}


\begin{acks}
We thank the National Science Foundation for its support under awards CNS-2234400 and CNS-2234401. Any opinions, findings, and conclusions or recommendations expressed in this material are those of the authors and do not necessarily reflect the views of the National Science Foundation.
\end{acks}

\bibliographystyle{ACM-Reference-Format}
\bibliography{sample-base}

\end{document}

%% file: sections/00_Abstract.tex
\begin{abstract}

Conducting a complete computer architecture simulation study
is challenging because configuration, execution, and
analysis are often encoded implicitly in scripts or directory
conventions rather than represented explicitly. As a result,
studies are difficult to scale, hard to reproduce, and dependent
on custom tooling at every stage.
We present \name{}, which makes the structure of a
simulation study explicit. In \name{}, hardware
topologies are described as declarative graphs that automatically
generate executable simulation code, eliminating hand-written
simulator programs. Stateless runners autonomously claim and
execute jobs from a shared experiment store, enabling
configuration-benchmark matrices to scale without manual
orchestration. Simulation outputs are stored as structured
artifacts tied to configurations, benchmarks, and hardware
components, enabling systematic result exploration without custom
parsers. We evaluate \name{} on a $12 \times 8 = 96$-configuration simulation
matrix spanning memory-bound, compute-bound, and mixed-intensity GPU
workloads. Declarative simulation specifications drive full simulations
with a median kernel time error of 0.18\% relative to hand-written
MGPUSim configurations across 95.8\% of configurations. The platform
introduces only 1.6 seconds of overhead per simulation, negligible
relative to realistic simulation workloads.

\end{abstract}

%% file: sections/01_Introduction.tex
\section{Introduction}

Simulation is central to computer architecture research. Building physical prototypes for each design is prohibitively expensive, so researchers rely on simulation to explore design choices and evaluate architectural ideas before hardware is built~\cite{akram2019survey}. Simulators such as gem5~\cite{binkert2011gem5, lowe2020gem5}, SST~\cite{rodrigues2011structural}, Multi2Sim~\cite{ubal2012multi2sim}, AccelSim~\cite{khairy2020accel} and MGPUSim~\cite{sun2019mgpusim} have become foundational infrastructure for the field. As architectural designs grow more complex, they incorporate heterogeneous accelerators, deep memory hierarchies, and increasingly large software stacks. As a result, simulation studies now explore larger design spaces and require executing thousands of simulations across many configurations and benchmarks.

\begin{figure}[t!]
  \centering
  \includegraphics[width=\linewidth]{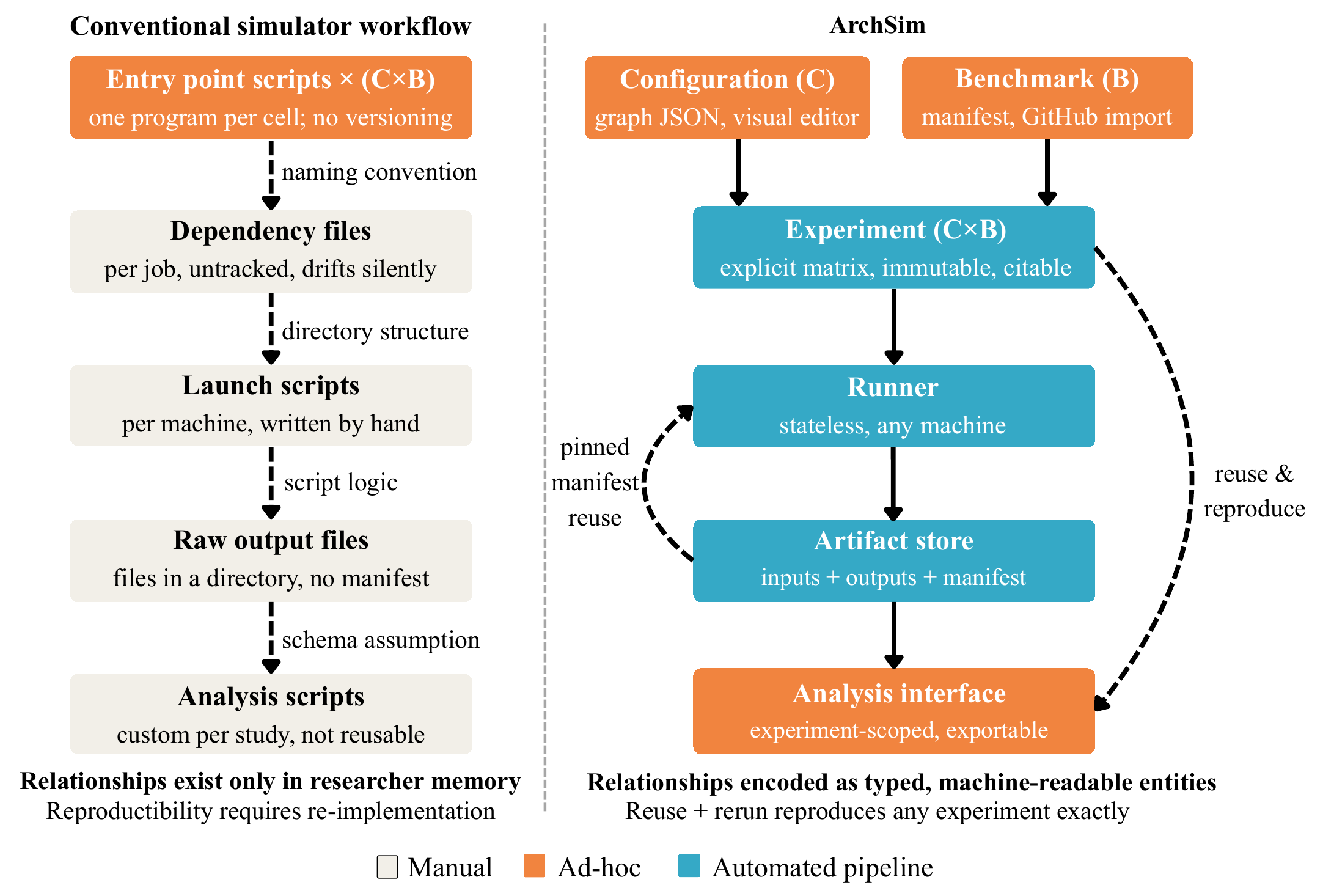}
  \caption{Conventional simulator workflow versus ArchSim.}
  \label{fig:parallel}
\end{figure}



Despite the maturity of individual simulators, the barrier to
conducting a complete simulation study remains high. In current
practice, researchers define hardware configurations as imperative
simulator source code rather than declarative specifications, orchestrate execution through hand-written, 
all tied together through naming conventions and directory
structures. This approach introduces four primary barriers:

\begin{itemize}[leftmargin=*, nosep]

\item \textbf{Configuration complexity.} Hardware topologies are
expressed as imperative simulator source code that instantiates
dozens of components, binds typed ports, and wires components
together in dependency order. Any change to the topology requires
manually updating every affected line across every configuration
variant, with no structural check that the result is consistent~\cite{binkert2011gem5, sun2019mgpusim}.

\item \textbf{Environment setup.} Executing a study across
multiple machines requires per-machine installation of the
correct toolchain, fetching of simulator and benchmark
repositories, and writing orchestration scripts that encode
the simulation structure in shell logic and directory naming,
with no portable description of the execution environment.

\item \textbf{Simulation extensibility and scalability.} Adding a new
configuration means updating every script, directory, and aggregation
pipeline that references the simulation structure by name. As the
number of configurations and benchmarks grows, this fragility
compounds: a study with $C$ configurations and $B$ benchmarks
requires maintaining $C \times B$ simulation programs with no
structured representation of the simulation as a whole. Because
relationships between configurations, benchmarks, and outputs are
encoded only in naming conventions and script logic, inconsistencies
are difficult to detect in practice~\cite{feitelson2015repeatability}.

\item \textbf{Result interpretability.} Analyzing results requires
custom parsing scripts tightly coupled to the simulator output
format and directory layout. Because simulation metrics are
spatially organized across hardware components, flat parsing
scripts cannot represent this structure, making it difficult to
trace a performance observation back to the specific component
that caused it~\cite{binkert2011gem5, sun2019mgpusim}.

\end{itemize}

Simulation study in computer architecture is not just a simulator run. It is a
structured scientific experiment consisting of hardware configurations,
benchmarks, a systematic execution of every configuration-benchmark
pair, and analysis that draws conclusions from the resulting data. The
four barriers above share a common root cause: the structure of the
experiment is encoded in code, scripts, and directory conventions
rather than being represented explicitly. When configuration is code,
it cannot be validated, stored, or re-executed without the original
developer's environment. When execution is orchestrated by scripts,
adding one configuration requires updating every artifact that encodes
the experiment structure. When outputs are files in a directory,
analysis requires custom parsers that break the moment the experiment
changes. Making the structure of a simulation experiment explicit and
machine-readable eliminates all four barriers systematically. Figure~\ref{fig:parallel} illustrates this contrast between conventional script-based workflows and \name{}’s explicit, structured representation of simulation experiments.


Prior work has addressed individual pieces of this problem.
Existing simulators provide detailed hardware models but leave
simulation structure encoded in the scripts and directory
conventions that surround them, with no platform-level support
for managing configurations, benchmarks, or results as structured
entities. Workflow management systems such as
MLflow~\cite{zaharia2018accelerating} and
ReproZip~\cite{chirigati2016reprozip} address reproducibility
for flat parameter spaces but do not represent hardware topology
graphs, typed port bindings, or configuration-benchmark matrices.

We introduce \name{}, a web-based simulation experiment platform
for computer architecture research that addresses each of the
four barriers directly. To eliminate configuration complexity,
\name{} provides a visual flow editor where researchers build
hardware topologies by placing component nodes on a canvas,
setting parameters, and connecting ports, with no simulator
source code written or maintained. \name{} automatically
generates a complete, correct simulation program from the graph.
To remove environment setup overhead, we introduce a runner that
handles toolchain setup, code generation, compilation, and
execution automatically on any registered machine. To scale
experiments without manual orchestration, \name{} materializes
the full $C \times B$ simulation matrix and queues all jobs
automatically, showing live per-cell status as simulations
complete. To make results interpretable without custom scripts,
\name{} provides comparative views of simulation results with
component-level drill-downs for deeper analysis.

\name{} aims to simplify and formalize the end-to-end workflow of architecture
simulation studies. This paper makes the following contributions:

\begin{itemize}[leftmargin=*, nosep]

\item \textbf{An explicit experiment model} that formalizes
    configurations, benchmarks, and experiments as first-class
    machine-readable entities, enabling experiments to be reproduced
    directly from their stored representation without reconstructing
    scripts or directory layouts.

\item \textbf{A declarative configuration representation and
    automatic code generation pipeline} that allows hardware topologies
    to be described as structured graphs rather than hand-written
    simulator programs, and automatically translates these descriptions
    into executable simulation code.

\item \textbf{A distributed execution model} in which
    stateless runners claim and execute simulation jobs autonomously
    from a shared experiment store, enabling large
    configuration-benchmark matrices to be explored without manual
    orchestration.

\item \textbf{A structured analysis interface} that records
    simulation outputs as structured artifacts tied to configurations,
    benchmarks, and hardware components, supporting systematic metric
    comparison and interactive result exploration without custom
    parsing scripts.

\end{itemize}


%% file: sections/02_Motivation.tex
\section{Design Requirements}\label{sec:motivation}

The artifacts that define an experiment are rarely captured
explicitly and often exist only in the researcher's mental model of
the experiment. Consider a simulation study that sweeps five cache
configurations across six benchmarks. A pattern of fragility
appears across architecture simulators: gem5 uses Python scripts to
instantiate and connect simulation objects, SST uses similar
programmatic configuration, and MGPUSim uses Go source programs. In each case, hardware
configurations are expressed as imperative code, so each
configuration variant requires a separate program or script.

We use MGPUSim to illustrate a concrete instance of this problem. A 5 $\times$ 6 experiment requires maintaining 30 separate simulation programs
along with benchmark inputs, output files, and analysis scripts linked
together through naming conventions and directory structures. As the
number of configurations and benchmarks grows, the resulting workflow
becomes increasingly fragile: small changes require editing many
programs, mistakes propagate silently across runs, and reproducing
results requires reconstructing the entire artifact graph.

\textbf{\underline{R-1}: Configuration and Code Generation.}
Defining the hardware system to simulate is the first and most
technically demanding stage of a simulation experiment. A typical
architecture topology may include cores, cache hierarchies, address
translation structures such as TLBs, memory controllers, and
interconnect components~\cite{binkert2011gem5,sun2019mgpusim,
khairy2020accel}. In current practice, topologies are expressed as
simulator source code, making component relationships invisible to
the user and impossible to validate without executing the program.
The most time-consuming errors in hand-written simulator workflows
are not logical mistakes but silent wiring faults: missing port
bindings that compile without error yet produce zero-metric outputs.
Identifying such faults requires reading binary output,
cross-referencing it with source, and manually tracing the component
graph across dozens of components and hundreds of port bindings.
Researchers should be able to define hardware
topologies through a structured visual interface that validates
wiring at definition time, and the system should automatically
generate correct executable simulation code from that definition,
eliminating hand-written simulator programs entirely.

\textbf{\underline{R-2}: Execution and Scalability.}
After a configuration is developed and verified, researchers evaluate
the design by running benchmarks against it. A study with $C$
configurations and $B$ benchmarks requires $C \times B$ compiled
simulation programs, each needing separate compilation, job scripts,
and result aggregation logic. Adding a new configuration requires
updating job scripts, file naming schemes, and aggregation pipelines
by hand, with no structured representation of the experiment as a
whole. Researchers should be able to easily run large
configuration-benchmark matrices across multiple machines, with
simulations scaling out automatically as more machines designs are added.

\textbf{\underline{R-3}: Reproducibility.}
Reproducing a simulation experiment in practice is rarely achieved.
The inputs required to reproduce a run, hardware configuration,
benchmark parameters, simulation seed, and metric extraction scripts, are scattered across simulator source files and ad hoc pipelines, making reproduction dependent on incomplete documentation or direct communication with the original authors~\cite{feitelson2015repeatability}.
Even when inputs are preserved, simulator version drift can silently
alter results between runs, yet the exact dependency versions used
are rarely recorded. Every simulation run should store its complete
input specification and resolved simulator dependency versions in a
structured, machine-readable format sufficient to reproduce the
run.

\textbf{\underline{R-4}: Result Analysis.}
Architecture simulation produces large volumes of performance data
organized across many hardware components. Understanding performance
requires comparing results across configurations and benchmarks while
also examining component-level behavior such as cache hit rates or
CPI stack breakdowns. In conventional workflows this analysis is
performed using custom scripts tightly coupled to the simulator
output format and the experiment directory structure. When the
experiment changes, these scripts must be updated manually, and
because the relationship between outputs and analysis logic is
encoded only in script logic, there is no mechanism to ensure
consistency. Simulation outputs should be stored in a structured,
queryable format alongside their input specifications, and researchers should be able to retrieve
experiment-level metrics and navigate to component-level details
for any selected configuration and benchmark without writing
custom parsing scripts.


%% file: sections/03_System_Overview.tex
\section{\name{}}\label{sec:features}

We first introduce the core concepts and entities that \name{} uses
to represent an experiment (Section~\ref{sec:org}), then
describe how each design decision satisfies the requirements identified in
Section~\ref{sec:motivation}. \name{} organizes the workflow of a
simulation study as a sequence of explicit stages spanning
configuration, execution, reproducibility, and analysis.

\subsection{Organization}\label{sec:org}


\name{} organizes the end-to-end workflow of an experiment study into
four areas: \textit{Model}, \textit{Configuration}, \textit{Experiment},
and \textit{Analysis} (see Figure~\ref{fig:simulation_run_complete}). Researchers first populate the Model library with
reusable hardware component definitions and benchmark workloads imported
from external repositories. They then use the Configuration editor to
assemble those components into complete hardware topologies, described
as declarative graphs and stored as JSON documents. Next, an Experiment
combines a set of configurations and benchmarks into a $C \times B$
simulation matrix, dispatches all jobs to registered runners, and
displays live per-cell status as simulations complete. Finally, the
Analysis view queries the resulting artifact store to compare metrics
across the matrix and drill down into component-level behavior for any
selected (configuration, benchmark) pair.

\begin{figure}[t!]
  \centering
  \includegraphics[width=\linewidth]{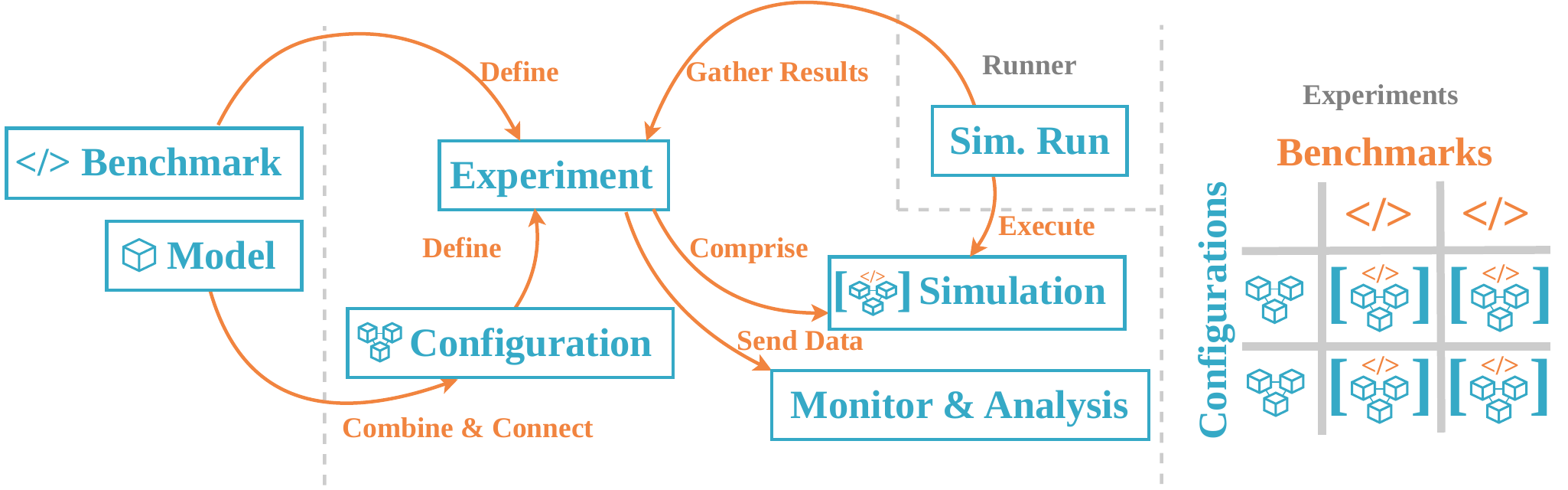}
  \caption{ArchSim experiment lifecycle. Researchers define benchmarks
  and configurations independently and combine them into an experiment,
  which materializes one simulation record per (configuration,
  benchmark) pair. Runners claim and execute simulations autonomously,
  reporting results to the artifact store. The Monitor \& Analysis
  interface queries the store directly, closing the loop from
  experiment definition to result interpretation without manual
  orchestration at any stage.}
  \label{fig:organization}
\end{figure}

Each area is built around a set of first-class entities that explicitly
capture the structure of an experiment. The Model area organizes
reusable component definitions and benchmark workloads as typed,
machine-readable objects. The Configuration
area represents hardware topologies as declarative graphs of model
instances, parameters, and port connections, and automatically
generates executable simulation code from them
(Section~\ref{sec:codegen}). The Experiment area manages the $C \times
B$ simulation matrix, dispatching jobs to stateless runners and
preserving complete input and version information for reproducibility
(Sections~\ref{sec:execution} and~\ref{sec:reproducibility}). The
Analysis area stores simulation outputs as structured artifacts tied to
configurations, benchmarks, and hardware components, supporting
systematic metric comparison without custom parsing scripts
(Section~\ref{sec:analysis}). Figure~\ref{fig:organization}
shows how these entities interact across the experiment lifecycle, and
the following subsections describe each in detail.

\textbf{Model.} A model encapsulates the behavior of a
single architectural component such as a cache, TLB, compute unit,
memory controller, or interconnect. We currently support Akita~\cite{akita-github} simulation engine given is high flexibility and simpler interface. We are working on supporting gem5~\cite{binkert2011gem5} and SST~\cite{rodrigues2011structural} with an adaptor using CGO~\cite{cgo}, a tool that bridges Go and C code.

\textbf{Benchmark.} A benchmark represents a workload that executes
on the simulated hardware. Like models, benchmarks are imported from
external repositories and described by a manifest that declares
the workload's parameters and builder package. A benchmark definition
is shared across all experiments and configurations that reference it,
so the same workload can be evaluated across many hardware
configurations without duplication.

\begin{figure}[t!]
  \centering
  \includegraphics[width=\linewidth]{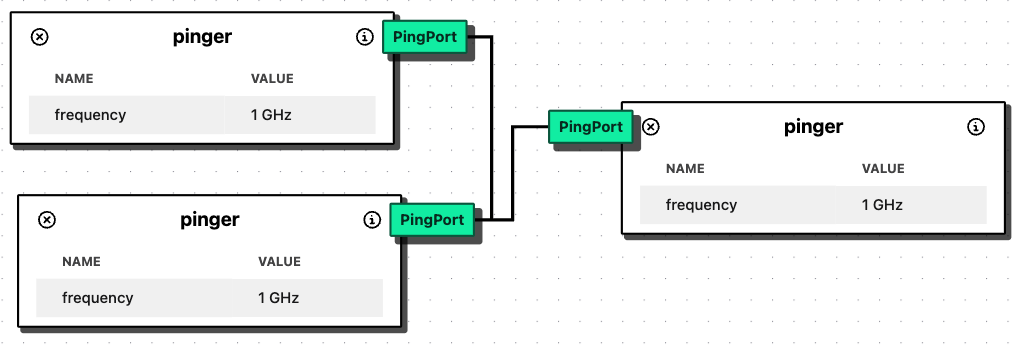}
  \caption{The ArchSim flow editor showing a three-component
  configuration. Each node represents a hardware model instance with
  an associated parameter panel; edges represent typed port connections
  between components.}
  \label{fig:pinger}
\end{figure}

\textbf{Configuration.} A configuration defines how hardware models
are instantiated, parameterized, and connected to form a complete
simulated hardware system. \name{} represents configurations as
graphs, where nodes correspond to hardware models and edges
represent interconnections between them (Section~\ref{sec:codegen}).
A flow editor provides a graphical interface for constructing and
modifying configurations without writing simulator code, as shown in
Figure~\ref{fig:pinger} (satisfying \underline{R-1}).



\textbf{Experiment.} An experiment represents a systematic evaluation
as an explicit $C \times B$ matrix of configurations and benchmarks.
To create an experiment, a researcher selects configurations and
benchmarks from their saved definitions. \name{} immediately renders
the full $C \times B$ grid, where each cell corresponds to one
(configuration, benchmark) pair to be simulated, as shown in Figure~\ref{fig:simulation_run_complete}. When the experiment
is launched, \name{} materializes one simulation record per cell. Each
record contains a complete, self-contained job specification including
the hardware configuration, benchmark parameters, simulation seed, and
trace settings, assembled from the current state of the selected
configurations and benchmarks at launch time. The specification is
stored immutably so that subsequent edits to a configuration or
benchmark do not silently alter queued or in-progress simulations.
The experiment grid reflects the live status of all simulations,
giving researchers a unified view of progress across the entire experiment matrix.


\textbf{Simulation.} A simulation corresponds to a
single $(\textit{configuration}, \textit{benchmark})$ pair in the
experiment matrix and represents the logical unit of work scheduled by
the system. Each simulation progresses through a lifecycle
\texttt{Ready} $\rightarrow$ \texttt{Running} $\rightarrow$
[\texttt{Done} | \texttt{Failed} | \texttt{Canceled}]. The resolved
simulator dependency versions are recorded after the first successful
run in an experiment and reused for all subsequent runs, allowing
\name{} to manage retries and track reproducibility metadata without
modifying the logical structure of the experiment.


\textbf{Runner.} A runner is any machine registered to an \name{}
account that can claim and execute simulation jobs on behalf of a
researcher. Once registered, a machine launches the
\texttt{archsim-runner} daemon via the CLI, which continuously
polls the backend for pending jobs, claims work autonomously,
executes the simulation pipeline, and reports results. Runners are classified as \emph{shared} or \emph{personal}.
Shared runners are available to all platform users and are
centrally managed by an institution or administrator. Personal
runners are registered by an individual researcher and are
accessible only to that user. Researchers can attach private
machines such as workstations or lab servers as personal runners
for their own simulations. Each runner polls the backend for
pending jobs, claims work autonomously, executes the simulation
pipeline, and reports results. All job-specific information is
included in the payload, so simulations can execute on runners
  without manual configuration (satisfying \underline{R-2}).
The runner is also responsible for translating the declarative
configuration JSON into an executable simulation program; this
code generation pipeline is described in detail in
Section~\ref{sec:codegen}.




\textbf{Analysis.} \name{} provides a structured analysis interface
that queries the artifact store generated during simulation execution.
At the experiment level, the system retrieves summary metrics across
all simulations from the Postgres database (satisfying \underline{R-4}).
At the simulation level, the interface supports component-level
drill-down for a selected $(\textit{benchmark}, \textit{configuration})$
pair (satisfying \underline{R-4}). Raw simulation artifacts may also
be stored in S3 for deeper post hoc analysis
(Section~\ref{sec:reproducibility}).

\subsection{Declarative Configuration and Code Generation}
\label{sec:codegen}

Architecture simulators typically construct hardware systems by
writing configuration files (e.g., Multi2Sim) or programs (e.g., gem5, MGPUSim) that instantiate components and
connect ports through builder APIs. This couples experiment
configuration with simulator source code, making it difficult to
generate and manage large families of hardware configurations for
systematic architectural studies.

\name{} addresses this by representing hardware topologies as
declarative configuration graphs that define the simulation to be generated (addressing \underline{R-1}).
The hardware structure is described as a structured document
specifying component instances, parameters, and interconnections,
allowing the system to automatically generate executable simulation
programs without researcher-authored code. 

\begin{figure}[t!]
 \centering
 \includegraphics[width=0.6\linewidth]{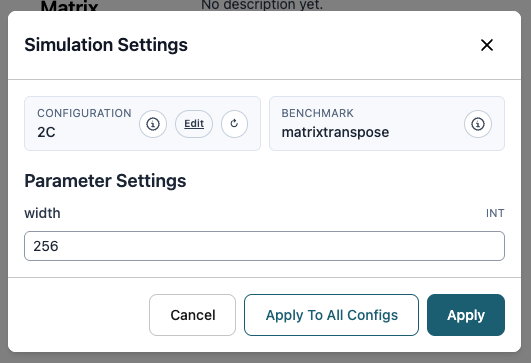}
 \caption{ArchSim parameter configuration dialog. Researchers
 adjust per-component parameters directly through the interface;
 changes are serialized into the configuration JSON and
 propagated to all affected simulations in the experiment matrix.}
 \label{fig:parameter_change_detail}
\end{figure}

\textbf{Configuration representation.}
Configurations are defined through a graphical flow editor where
researchers place component nodes, set parameters, and draw port
connections between components. The resulting topology is
serialized into a JSON document by \name{}. Nodes represent
component instances, each with a \texttt{name}, a
\texttt{builder\_package} identifying the builder to invoke, and
a \texttt{params} list of parameter assignments. A builder is a
factory module whose construction API instantiates a component
class; it is identified by its import path, which the code
generation pipeline uses to invoke the correct constructor when
assembling the executable simulation program. Parameter values
may be literals, variable references resolved during code
generation, or structured port bindings referencing another
component and port. Edges carry a \texttt{name} and a list of
\texttt{plugs} identifying the component-port pairs to connect.
Together, nodes and edges form a directed graph that is validated
before code generation: the runner checks that all port bindings
and plugs reference existing components, verifies that referenced
variables are defined, and confirms that the dependency graph
contains no cycles. These checks catch structural errors such as
dangling references and circular dependencies at definition time
rather than at compile time or simulation runtime. Benchmark-level
parameters (Figure~\ref{fig:parameter_change_detail}), are set
separately at experiment launch time and serialized into the simulation job when the researcher clicks \texttt{Run Simulations}, as shown in Figure~\ref{fig:simulation_run_complete}.

\textbf{Configuration space.} The configuration abstraction supports
four categories of architectural parameters, all expressible as
declarative JSON edits without modifying simulator source code:
component topology (adding, removing, or rewiring components in the
graph), memory hierarchy (cache capacity, associativity, and block
size), compute unit microarchitecture (pipeline width and subsystem
parameters), and address translation and memory control (TLB
structure and memory controller arrangement).
Section~\ref{sec:casestudy} demonstrates this directly across all
four categories.




\textbf{Code generation pipeline.}
Given a configuration JSON and a benchmark manifest, the runner
generates a complete simulation program, compiles it, and executes
it automatically, producing structured performance metrics and
optional trace artifacts for tools such as
Daisen~\cite{sun2021daisen} (satisfying \underline{R-1}).
\name{} does not modify the simulator core; instead, the runner
synthesizes a self-contained host program that imports the existing
simulator component libraries and benchmark packages and invokes
their public builder APIs. The only requirement from the simulator
libraries is the stable builder-style construction APIs they already
expose.

The generation pipeline proceeds in five steps. First, the
configuration JSON is parsed into an internal graph representation.
Second, the graph is validated for structural correctness, checking
component references, variable definitions, and dependency cycles.
Third, the runner emits a Go source file containing the simulation
engine initialization, topologically ordered component builder
calls, port connection wiring, benchmark setup, and optional trace
instrumentation. Fourth, module dependencies are resolved using the
experiment's pinned version manifest. Fifth, the program is compiled
and executed, and the resulting metrics and artifacts are uploaded
to the artifact store.

Because these translation steps follow a fixed set of rules, the
generated program is a deterministic function of the configuration
JSON and benchmark parameters, so storing the configuration is
sufficient to reproduce the exact simulation program used in any
experiment. The same pipeline applies regardless of whether edits
change only parameter values or also add and remove nodes and edges.

\subsection{Pull-Based Distributed Execution}
\label{sec:execution}

\name{} addresses \underline{R-2} through a
pull-based distributed model in which stateless runners claim
self-contained jobs from a shared backend queue. Because the job
payload is complete, containing the configuration JSON, benchmark
parameters, simulation seed, and pinned version manifest, any
machine running the \texttt{archsim-runner} CLI can join the
execution pool and begin executing simulations without prior
simulator installation or manual environment configuration.

\begin{figure}[t!]
 \centering
 \begin{subfigure}[t]{\linewidth}
  \centering
  \includegraphics[width=0.8\linewidth]{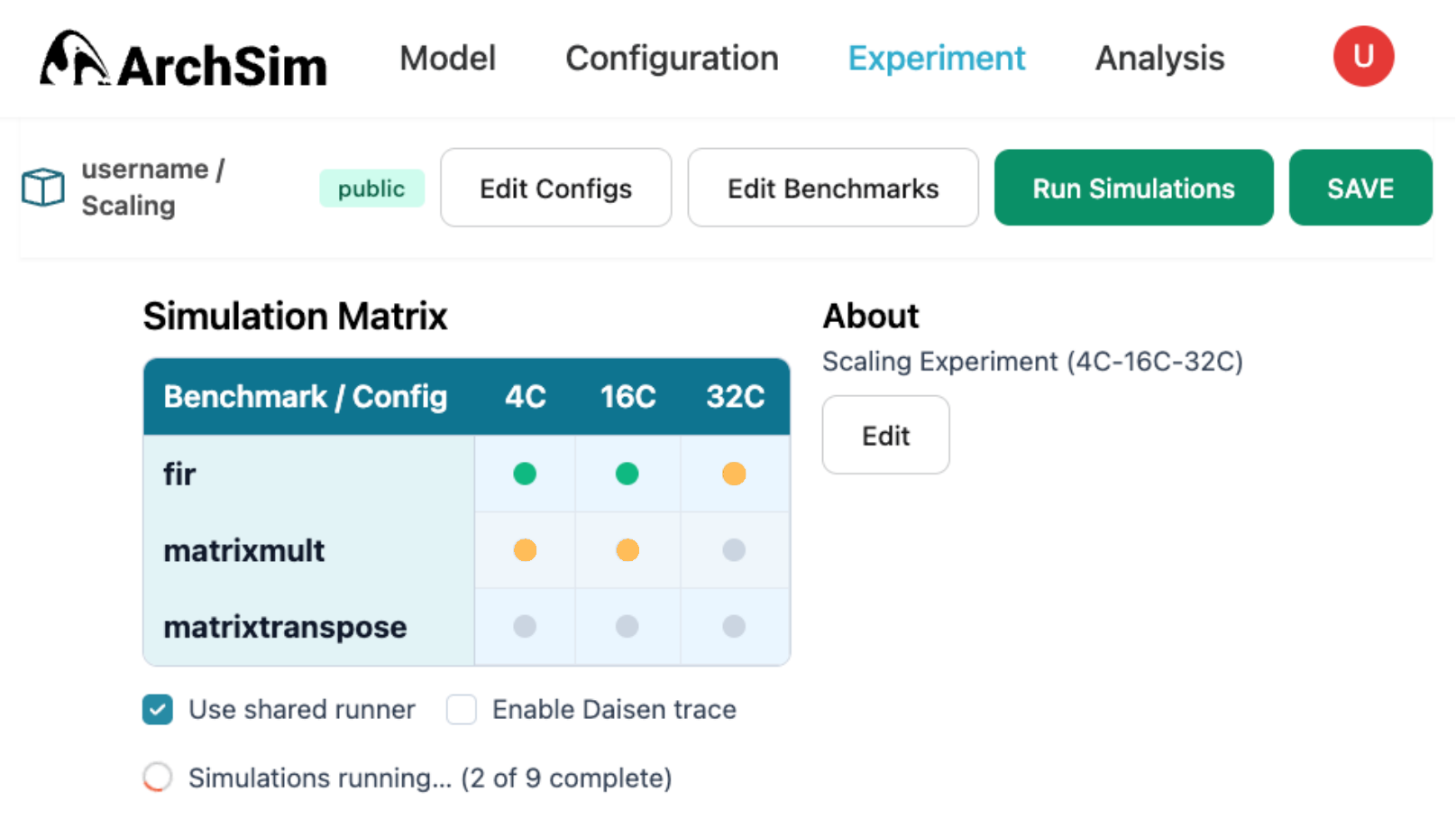}
  \caption{The simulation
 matrix during execution, showing per-cell status indicators
 for a 3-configuration $\times$ 3-benchmark study. Configurations
 are ordered by CU count (4C, 16C, 32C) along the columns and
 benchmarks along the rows.}
  \label{fig:simulation_run}
 \end{subfigure}

 \vspace{0.4em}

 \begin{subfigure}[t]{\linewidth}
  \centering
  \includegraphics[width=0.8\linewidth]{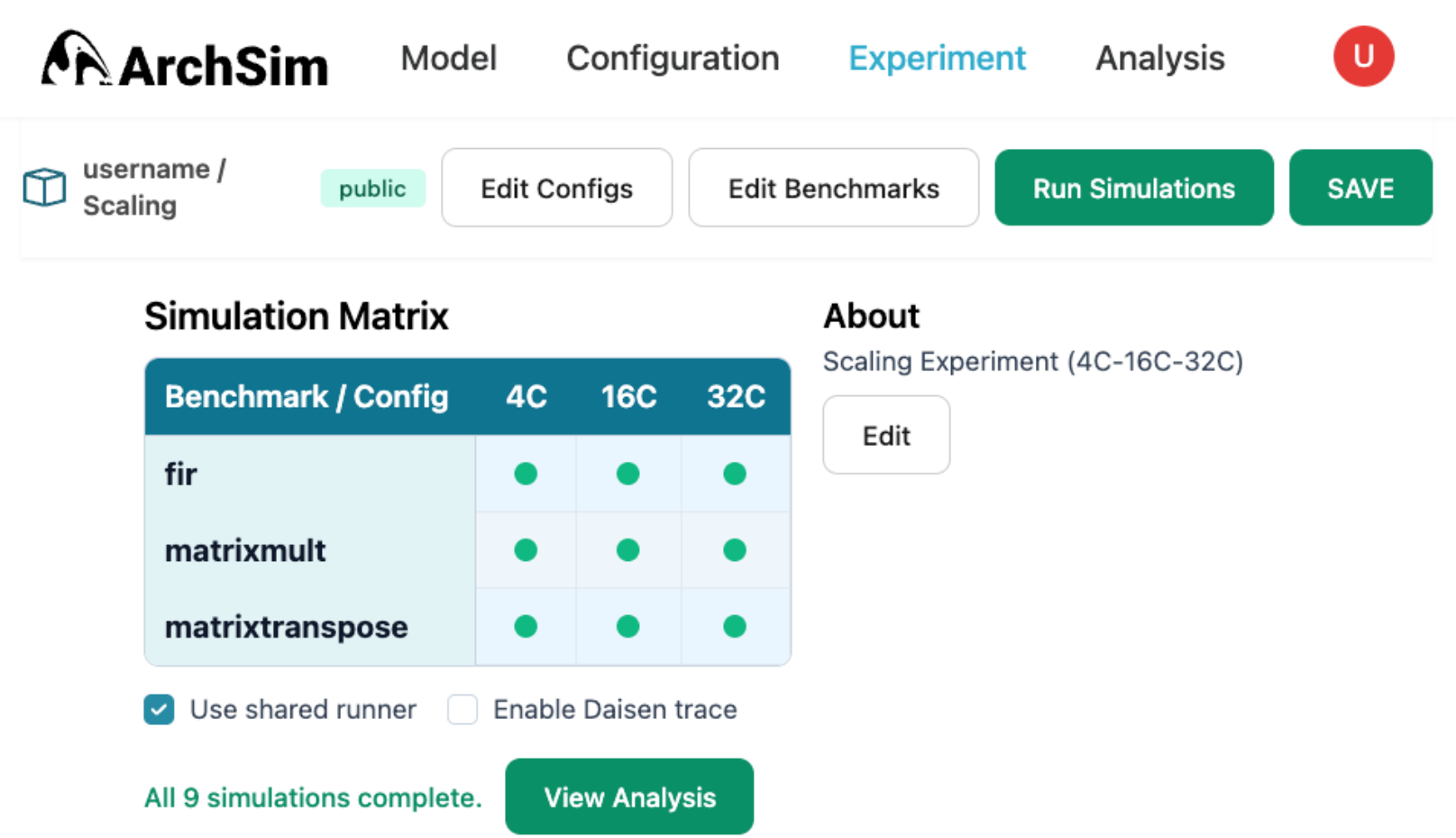}
  \caption{The same dashboard after all 9
 simulations have completed, with the View Analysis button
 enabled. Researchers can immediately identify failed cells and re-run the experiment after correcting errors without navigating away from the experiment view.}
  \label{fig:parameter_change}
 \end{subfigure}
 \caption{ArchSim experiment dashboard.}
 \label{fig:simulation_run_complete}
\end{figure}

\textbf{Job lifecycle.}
When a researcher clicks \texttt{Run Simulations}, as shown in Figure~\ref{fig:simulation_run_complete}, the backend materializes
one simulation record per cell in the $C \times B$ matrix, each
entering the \texttt{Ready} state. By default, re-running an experiment resets incomplete and failed cells to \texttt{Ready} with refreshed payloads; an incremental mode is also available that skips cells already marked \texttt{Done}, allowing researchers to fill in only missing results.

\textbf{Runner execution.}
Runners periodically poll the backend for available jobs and
claim work independently. Once a job is claimed, the runner
generates the simulation program, compiles it, executes the
benchmark, reports the result, and uploads generated artifacts. Claim rules are enforced server-side, and jobs that remain \texttt{Running}
beyond a timeout are reset to \texttt{Ready} and returned
to the queue. The experiment dashboard
(Figure~\ref{fig:simulation_run_complete}) displays the full
$C \times B$ matrix with per-cell outcomes, allowing researchers
to identify failed pairs and resubmit after correcting errors.


\subsection{Reproducibility through Explicit Artifact Preservation}
\label{sec:reproducibility}

\name{} ensures experiment reproducibility by explicitly preserving
both the inputs and outputs of every simulation run. Instead of
treating simulation programs, configuration parameters, and results
as loosely connected files in a researcher’s working directory,
\name{} records them as structured artifacts associated with each
experiment. This design allows simulations to be reconstructed and
re-executed directly from stored records on any machine running the \texttt{archsim-runner} CLI, independent of the environment that originally queued the job.


Each queued simulation stores its complete job specification as a single JSON document containing both the hardware configuration and benchmark sections. The platform additionally persists relational identifiers (experiment, configuration, benchmark), run status, structured metrics, and optional large binary artifacts via storage URLs.

\textbf{Complete input preservation.}
The platform retains the full input specification for each run; when a simulation completes,
structured outputs are stored alongside it (addressing \underline{R-3}). The stored record includes the hardware configuration document, benchmark definition
and runtime parameters, and the simulation seed. Together these inputs are sufficient to regenerate the simulation program and re-execute the experiment. Because \name{}’s code generation pipeline is deterministic, reconstructing a simulation from the stored configuration produces a program that is semantically equivalent to the original run for a given runner version.

\textbf{Experiment-level version manifest.}
Preserving configuration inputs alone is insufficient if the underlying
simulator evolves over time. \name{} therefore records the exact
simulator dependency versions used during execution (addressing
\underline{R-3}). When the first simulation in an experiment completes
successfully, the resolved Go module graph (\texttt{go.mod} and
\texttt{go.sum}) is captured and stored as the version manifest for
that experiment. All subsequent simulations in the experiment reuse
this manifest when claiming jobs, ensuring that every run executes
against the same simulator versions even if upstream repositories
change. This mechanism is automatic and requires no manual version
management by the researcher. Pinning versions at the experiment level
ensures that all cells in the $C \times B$ matrix execute against the
same simulator build, so differences in results reflect only
configuration and benchmark inputs rather than toolchain variation.

Different experiments may
use different simulator versions depending on when their first
simulation completed. If no simulation in an experiment completes
successfully before all runs fail, the manifest is not established
and version pinning does not take effect for that experiment.

\textbf{Structured output storage.}
Simulation outputs are stored in a relational schema: each metric
row is keyed to the experiment, configuration, and benchmark that
produced it, and each metric names the simulated hardware component
it refers to using a hierarchical component path in the simulator's
naming scheme (addressing \underline{R-3}). This enables
systematic comparison across the experiment matrix without custom
parsing scripts, since results can be grouped by 
configuration or benchmark, directly
from the store. Raw simulation artifacts may also be stored for
deeper post hoc analysis when needed.




\subsection{Result Analysis}
\label{sec:analysis}

\name{} provides a result analysis feature that operates directly on the
structured artifact store produced during simulation execution.
Because each metric is associated with its experiment and (configuration, benchmark) cell, and tagged with a component path in the simulators's metric schema, the interface supports both simulation comparisons and per-component inspection without custom parsing scripts. The interface renders a bar chart of a selected metric
(e.g., \texttt{kernel\_time}) across all $C \times B$ pairs,
supporting grouping by configuration or benchmark to reflect
the two standard axes of comparison in simulation studies:
architectural design alternatives and workload characteristics. Users can drill down
through the component hierarchy to inspect component-level metrics
and distributional summaries for any selected (configuration,
benchmark) pair. For reproducible figures, the interface exports
standalone Python scripts with embedded tabular data, allowing
plots to be regenerated and restyled for publication. Analysis
beyond what the structured artifact store exposes, such as
fine-grained execution traces, is supported through integration
with tools such as Daisen~\cite{sun2021daisen}.


\textbf{Experiment-scoped metric retrieval.}
The interface retrieves all simulation runs belonging to a selected
experiment and renders a bar chart comparison view across the
$C \times B$ configuration-benchmark matrix for a selected metric, for example, simulated execution time (addressing \underline{R-4}). Each bar corresponds to a stored metric
value for a specific \texttt{(configuration, benchmark)} pair, letting
researchers visually identify best- and worst-performing configurations
at a glance. Results can be grouped by configuration or by benchmark,
enabling quick comparison of architectural alternatives or workload
behaviors across the experiment without intermediate aggregation
pipelines, as shown in Figure ~\ref{fig:case_overview}.


\textbf{Component-level drill-down.}
Selecting a specific \texttt{(configuration, benchmark)} pair opens a
component-level view derived from the stored metric locations
(addressing \underline{R-4}). Components are organized hierarchically
according to the simulator's hardware structure, enabling navigation
from experiment-level summaries to measurements collected at
individual components such as compute units, caches, or memory
controllers (Figure~\ref{fig:case_drilldown}). For example, a researcher can inspect L2 read and write
hit rates, L3 MSHR activity, or per-CU CPI stack contributions
directly from this view, for any component that emits these counters,
without writing any parsing scripts. Section~\ref{sec:casestudy}
demonstrates this workflow concretely: an anomaly identified in the
experiment-level bar chart is traced to near-zero L2 hit rates and
nonzero L3 MSHR activity through the component-level drill-down.



%% file: sections/04_Evaluation.tex
\section{Evaluation}\label{sec:eval}

We evaluate ArchSim across four studies: simulation accuracy (Section~\ref{sec:eval:accuracy}), which establishes that ArchSim-generated simulations are consistent with hand-written MGPUSim configurations; reproducibility (Section~\ref{sec:eval:repro}), which establishes that ArchSim introduces no additional non-determinism beyond the underlying simulator; runner scalability (Section~\ref{sec:eval:scalability}), which evaluates the speedup delivered by the pull-based execution model as the number of concurrent runners grows; and execution overhead (Section~\ref{sec:eval:overhead}), which characterizes the cost of ArchSim's code generation and compilation pipeline relative to a pre-compiled baseline.

\subsection{Methodology}\label{sec:eval:methodology}

\textbf{Platform.}
Our experiments are conducted on a dual-socket AMD EPYC
7543 machine (64 cores, 128 hardware threads) with 251 GiB of main
memory, running Ubuntu 22.04.5 LTS (kernel 5.15.0-173) and Go
1.26.1. All simulations are compiled and executed by ArchSim runners
registered on this machine. \name{}'s record-and-reuse version manifest ensures that every simulation in each experiment uses
identical \texttt{akita} and \texttt{mgpusim} dependency
versions.

\textbf{Simulation configuration.}
We integrate \name{} with the MGPUSim simulator built on the
Akita~\cite{akita-github} simulation engine. Hardware configurations
are defined declaratively in \name{}'s flow editor as directed
graphs of MGPUSim components. We evaluate eight CU configurations
(1, 2, 4, 8, 16, 32, 64, 128 CUs) with all other topology
parameters fixed. The CU sweep is chosen as the primary dimension because it exercises the code generation pipeline across a wide range of topology scales, from a single compute unit with a handful of components to a 128-CU system with hundreds of port bindings. The ArchSim configuration interface is not limited to this dimension: any parameter exposed by the MGPUSim builder API, including cache capacity, associativity, block size, interconnect bandwidth, TLB size, and memory controller count, can be modified through the flow editor's parameter panel or swept across configurations without touching simulator source code. The case study in Section 5 demonstrates this directly: the four cache hierarchy configurations differ not only in the number of cache levels but also in associativity (4-way vs. 16-way), block size (64B vs. 128B), and the structural presence or absence of the L1.5 and L3 components, all expressed as declarative JSON changes. Each configuration is defined once in the flow
editor and reused across all benchmarks and evaluation studies
without modification.

\textbf{Benchmarks.}
To ensure the robustness of our evaluation, we use 12
representative benchmarks drawn from five suites:
AMDAPPSDK~\cite{staff2014opencl},
HeteroMark~\cite{sun2016hetero},
PolyBench~\cite{polybench},
DNNMark~\cite{dong2017dnnmark}, and
Rodinia~\cite{che2009rodinia}, covering the full range
of GPU workload characteristics.


Benchmark parameters are listed in Table~\ref{tab:benchmarks} 
and are held constant across both the accuracy and reproducibility 
studies.

\begin{table}[h]
\centering
\caption{Benchmark parameters used across all evaluation studies.}
\label{tab:benchmarks}
\begin{tabular}{lll}
\toprule
Abbrev. & Benchmark & Parameters \\
\midrule
ATX  & atax               & $512 \times 512$ \\
BiCG & bicg               & $512 \times 512$ \\
BTS  & bitonicsort        & length $= 32{,}768$ \\
FWT  & fastwalshtransform & length $= 16{,}384$ \\
FIR  & fir                & length $= 262{,}144$ \\
FW   & floydwarshall      & 512 nodes, 1 iteration \\
MM   & matrixmult         & $128 \times 128 \times 128$ \\
MT   & matrixtranspose    & width $= 2{,}048$ \\
NB   & nbody              & 4{,}096 particles, 3 iterations \\
NW   & nw                 & length $= 1{,}024$, penalty $= 10$ \\
ReLU & relu               & length $= 4{,}194{,}304$ \\
SC   & simpleconvolution  & $512 \times 512$, mask $= 3$ \\
\bottomrule
\end{tabular}
\end{table}

Together, the 12 benchmarks and 8 CU configurations produce a
$12 \times 8 = 96$-configuration experiment matrix that serves
as the common evaluation substrate for simulation accuracy and reproducibility studies below.

\subsection{Simulation Accuracy}\label{sec:eval:accuracy}

We evaluate whether \name{}-generated simulations produce results
consistent with a reference MGPUSim implementation of the same
platform topology. The reference implementation was
written by hand using MGPUSim's builder APIs directly, following
the same process a researcher would use without \name{}, and serves
as the ground truth against which \name{}'s code generation
accuracy is measured. Both runners use identical component parameters,
benchmark inputs, and simulation seed (seed~$= 0$) under a serial
discrete-event engine, eliminating all sources of non-determinism.

\begin{figure}[t]
    \centering
    \includegraphics[width=\columnwidth]{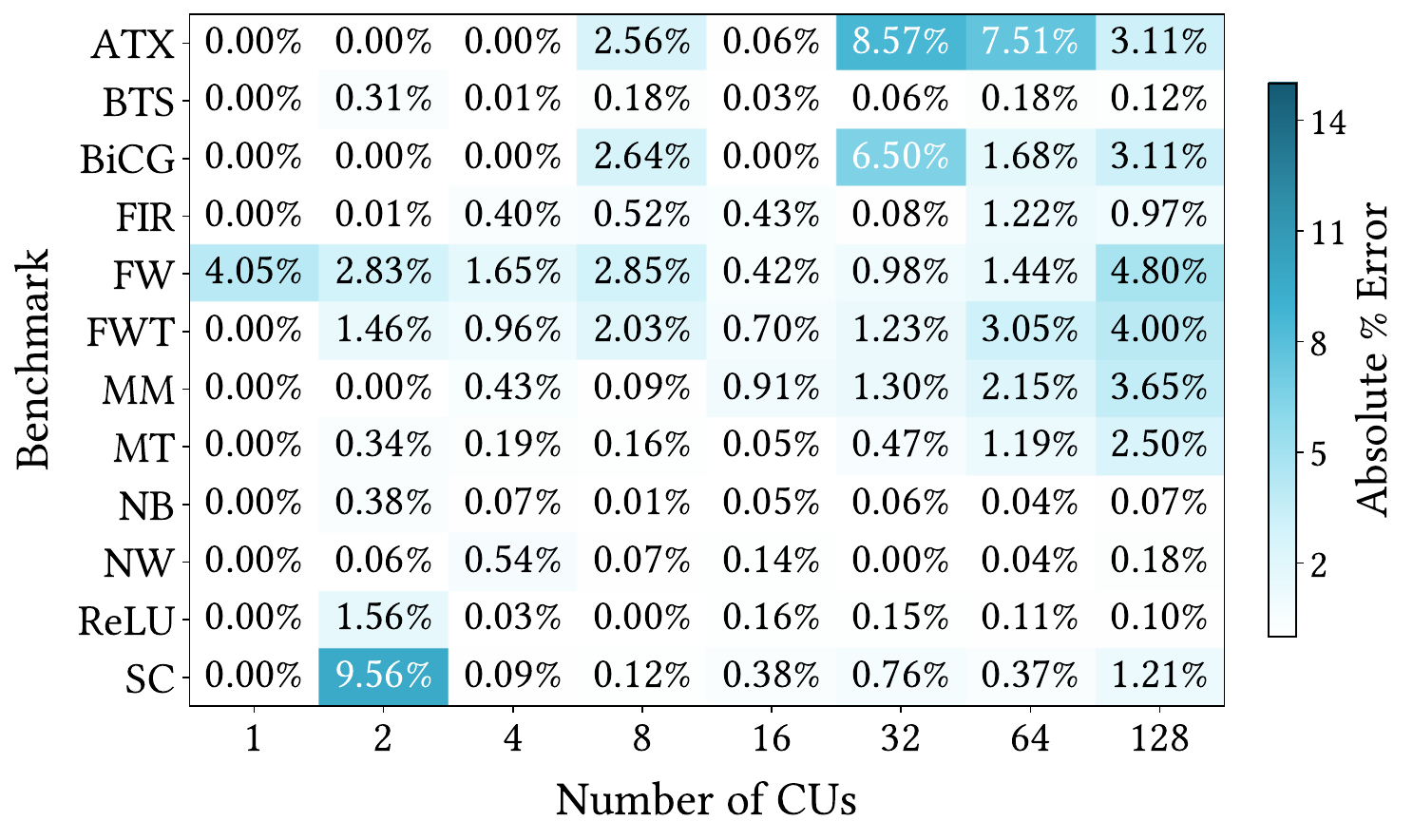}
    \caption{Absolute percentage error of \texttt{kernel\_time}
    between \name{} and the MGPUSim baseline across 12 benchmarks
    and 8 CU configurations. 95.8\% of configurations achieve
    RE $<$ 5\%. Darker cells are concentrated in the sub-millisecond
    execution range, where small absolute differences magnify into
    large percentage errors. 
    }
    \label{fig:accuracy_heatmap}
\end{figure}

Figure~\ref{fig:accuracy_heatmap} shows absolute relative error
(RE) across all 96 configurations and Table~\ref{tab:accuracy_summary}
summarizes the results. 95.8\% of configurations achieve RE $<$ 5\%
and the median RE of 0.18\% confirms the distribution is heavily
concentrated near zero. Mean RE remains below 2\% across all CU
counts, reaching its lowest at 16 CUs (0.28\%) and highest at 128
CUs (1.98\%), indicating that the code generation pipeline handles
structural translation correctly across topologies ranging from a
single CU to 128 CUs.

\begin{table}[h]
\centering
\caption{Simulation accuracy summary across 96 (benchmark, CU
count) configurations.}
\label{tab:accuracy_summary}
\begin{tabular}{lr}
\toprule
\textbf{Metric} & \textbf{Value} \\
\midrule
Total configurations evaluated  & 96              \\
Exact matches (RE $<$ 0.005\%)  & 19 \ (19.8\%)   \\
Within 0.5\% RE                 & 60 \ (62.5\%)   \\
Within 1\% RE                   & 68 \ (70.8\%)   \\
Within 5\% RE                   & 92 \ (95.8\%)   \\
Mean RE                         & 1.07\%          \\
Median RE                       & 0.18\%          \\
Maximum RE                      & 9.56\%          \\
\bottomrule
\end{tabular}
\end{table}

Accuracy varies by workload characteristic. The most accurate
benchmarks are NB (mean RE 0.08\%), BTS (0.11\%), and NW (0.13\%),
whose execution is dominated by compute or regular memory access
patterns. The least accurate are ATX (mean RE 2.73\%), FW (2.38\%),
and BiCG (1.74\%), which have irregular memory access patterns that
produce cache-sensitive timing differences at higher CU counts. The
four configurations exceeding 5\% RE (SC at 2 CUs, ATAX at 32 and
64 CUs, BiCG at 32 CUs) all involve sub-millisecond absolute
execution times where small absolute differences produce
disproportionately large percentage errors; the absolute differences
are architecturally insignificant.

\subsection{Reproducibility}\label{sec:eval:repro}

Simulator codebases evolve continuously, and even a minor internal
change to a dependency can silently alter simulation outputs without
any public API change or user warning. Without explicit version
pinning, two researchers running the same experiment may silently
execute against different simulator versions. \name{} addresses this
by recording a version manifest after the first successful run and
reusing it for all subsequent runs in the experiment.

\textbf{Version pinning is necessary and sufficient.}
We run the full 96-configuration simulation matrix under two
conditions. Trial A uses \name{}'s default pinned manifest, fixing
all dependencies to exact versions (\texttt{mgpusim v4.2.0},
\texttt{akita v4.9.2}). Trial B replaces \texttt{mgpusim} with a
fork containing a single one-line change: doubling the vector memory
transaction pipeline width in the compute unit builder. This change
modifies no public API and would be invisible to a user pulling the
latest dependency version. Under Trial A, zero mismatches are observed
across three trials and 288 total runs. Under Trial B, 8 of 96
configurations crash with verification failures and 66 of the
remaining 88 show measurable drift with a mean absolute deviation of
1.08\% and a maximum of 14.66\% (bicg at 16 CUs), with no
user-visible warning. Version pinning is therefore both necessary and
sufficient for reproducible simulation.

\begin{figure}[t]
    \centering
    \includegraphics[width=\columnwidth]{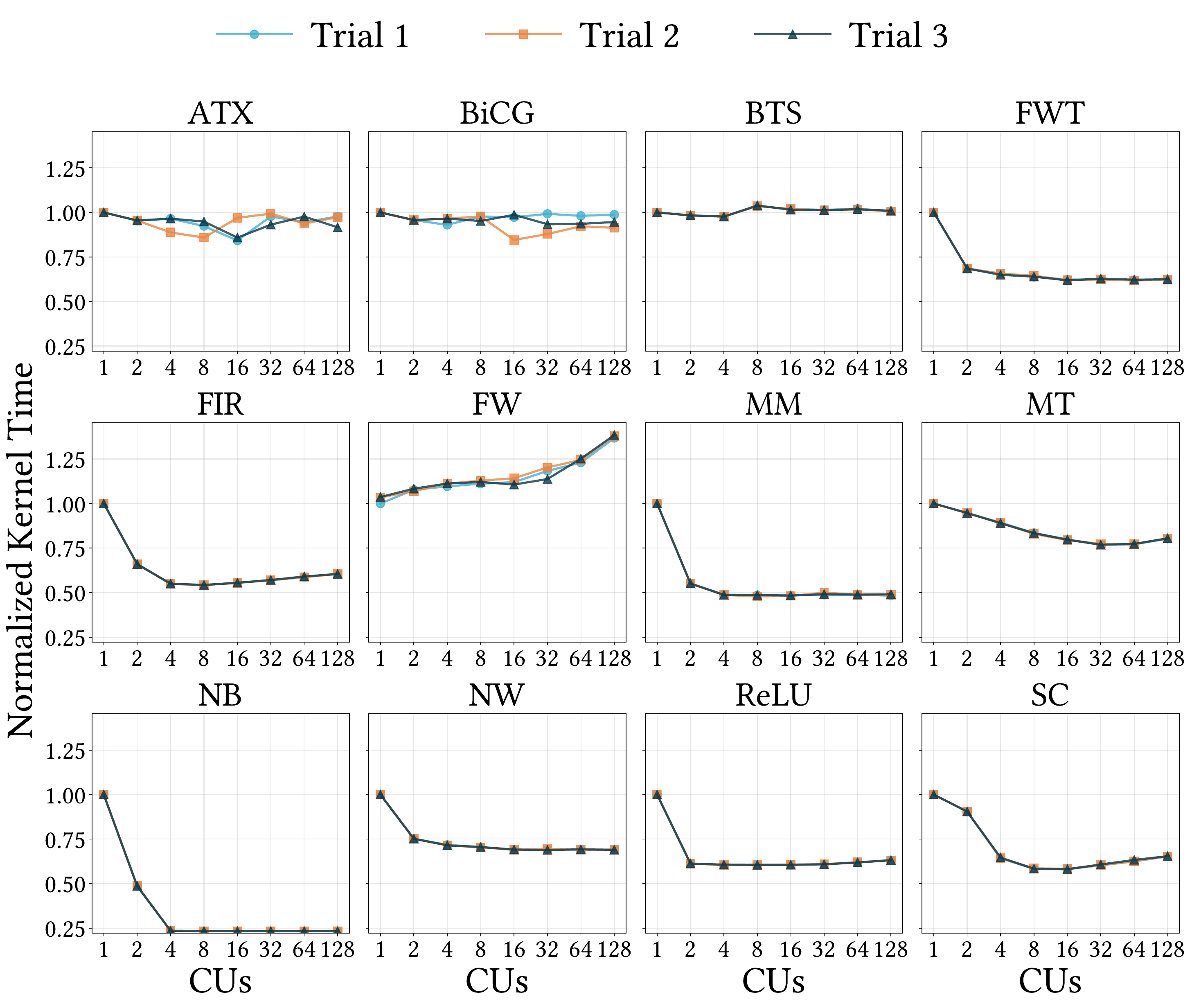}
    \caption{Normalized \texttt{kernel\_time} across three
    independent \name{} trials for twelve benchmarks (1--128 CUs).
    Overlapping lines at low CU counts confirm bit-identical
    results. Residual separation at high CU counts reflects
    simulator-level parallel event ordering, not platform
    non-determinism.}
    \label{fig:repro_overlay}
\end{figure}

We next verify that \name{} itself introduces no additional
non-determinism beyond the underlying simulator when versions are
pinned. We execute each of the 96 configurations three times under
the same pinned version manifest, configuration JSON, benchmark
parameters, and seed. Figure~\ref{fig:repro_overlay} shows normalized
\texttt{kernel\_time} across three trials for twelve benchmarks
spanning all CU counts.

\textbf{\name{} introduces no platform non-determinism.}
At 1 and 2 CUs, where the simulator runs serially, 9 of 12
benchmarks (75\%) produce bit-identical \texttt{kernel\_time}
values across all three trials (CV = 0.00\%). The exceptions show variation attributable to floating-point operation ordering in their kernels, a simulator-level property independent of ArchSim. Any variation
at higher CU counts is attributable entirely to the simulator's
parallel event ordering. Mean CV grows from 0.04\% at 2 CUs to
1.31\% at 16 CUs, consistent with parallel discrete-event
simulation behavior present in any MGPUSim runner. Overall,
83.3\% of configurations achieve CV $<$ 1\% and 95.8\% achieve
CV $<$ 5\%. Averaging across three trials, mean RE against the
baseline is 1.13\% (median 0.30\%), consistent with the
single-run result from Section~\ref{sec:eval:accuracy}, confirming
that inter-trial variation is random noise with no directional
component.

\subsection{Runner Scalability}\label{sec:eval:scalability}

\name{}'s pull-based execution model is designed to scale
horizontally by adding runner processes that claim jobs
autonomously from the shared queue without any centralized
coordination. We evaluate whether this design delivers
practical speedup as the number of concurrent runners grows.

We deploy 1, 2, 4, and 8 concurrent runner processes on the
same host, each executing the \texttt{archsim-runner} binary
independently. Runners poll the backend for pending jobs,
claim work via a first-writer-wins database update, and report
results without communicating with each other. The job queue
consists of 24 simulations (6 benchmarks $\times$ 4 CU
configurations: 1, 4, 16, and 64 CUs). We measure wall-clock time from the first job being claimed to
the last job reaching \texttt{Done} status and report speedup
and parallel efficiency relative to the single-runner baseline.

\begin{figure}[t]
    \centering
    \includegraphics[width=0.7\columnwidth]{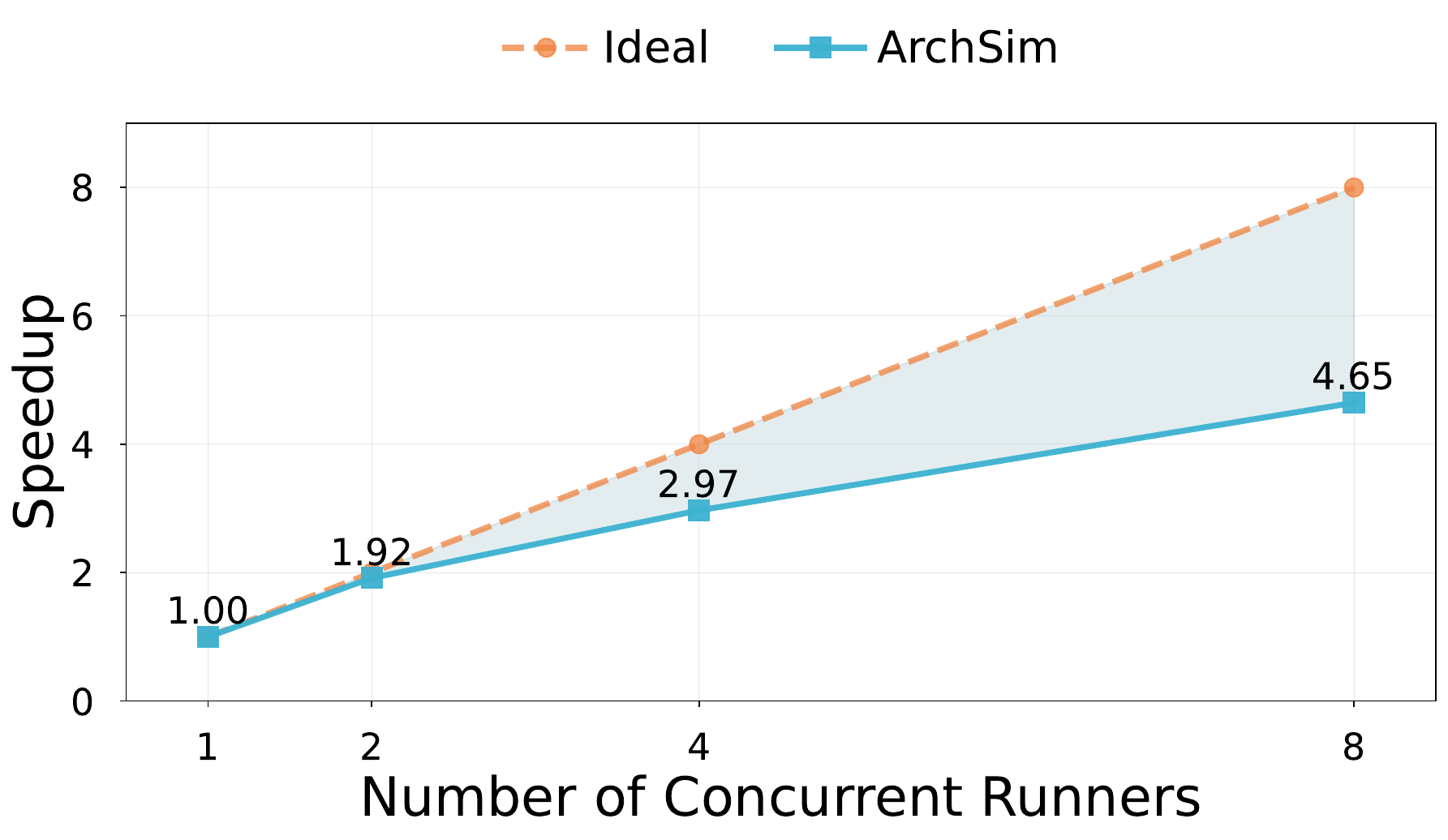}
    \caption{Speedup of \name{} with 1--8 concurrent runners
    against the ideal linear baseline. \name{} achieves
    near-linear scaling up to 4 runners and 4.65$\times$
    speedup at 8 runners, reducing total experiment wall-clock
    time from 36 minutes to under 8 minutes.}
    \label{fig:speedup}
\end{figure}

Figure~\ref{fig:speedup} shows the speedup curve against the ideal
linear baseline, while Table~\ref{tab:scalability} reports the
efficiency trend across all runner configurations.

\textbf{Speedup and wall-clock reduction.}
Table~\ref{tab:scalability} summarizes the results. With 2
runners, \name{} achieves 1.92$\times$ speedup at 95.8\%
efficiency, confirming that the pull-based job claiming
mechanism introduces negligible coordination overhead at low
concurrency. With 4 runners, speedup reaches 2.97$\times$
(74.3\% efficiency), approaching the ideal 4$\times$ and
remaining well above the 70\% threshold commonly used to
characterize well-optimized parallel systems. With 8 runners,
speedup reaches 4.65$\times$, reducing total experiment
wall-clock time from 36 minutes 12 seconds to 7 minutes 47
seconds, an 81\% reduction. All 24 simulations complete
successfully in every runner configuration with zero
contention-related failures.

\begin{table}[t]
\centering
\caption{Runner scalability results across 24 simulations.
Efficiency = actual speedup / number of runners. 
}
\label{tab:scalability}
\begin{tabular}{rrrr}
\toprule
\textbf{Runners} & \textbf{Wall-clock} &
\textbf{Speedup} & \textbf{Efficiency} \\
\midrule
1 & 36m 12s & 1.00$\times$ & 100.0\% \\
2 & 18m 53s & 1.92$\times$ & 95.8\%  \\
4 & 12m 11s & 2.97$\times$ & 74.3\%  \\
8 &  7m 47s & 4.65$\times$ & 58.1\%  \\
\bottomrule
\end{tabular}
\end{table}

\textbf{Efficiency degradation is host-level, not
architectural.}
The efficiency drop from 95.8\% at 2 runners to 58.1\% at 8
runners is consistent with shared CPU and I/O resource
contention on a single host as all runner processes compile
and execute simulations concurrently. This contention is
not a property of \name{}'s pull-based coordination model,
which is confirmed by the job isolation analysis: the median
coefficient of variation of individual job execution times
across all runner configurations is 0.89\%, indicating that
runners do not interfere with each other's simulation
execution. The efficiency degradation at 8 runners reflects
the host's compute capacity rather than any bottleneck in the
job claiming, artifact storage, or version manifest
mechanisms. Deploying runners across multiple hosts would
recover linear scaling beyond the single-host limit.

\textbf{Load distribution is balanced.}
The jobs are distributed evenly across all runner
configurations: 2 runners claim 12 jobs each, 4 runners
claim 6 jobs each, and 8 runners claim 3 jobs each. No
runner starvation or overloading is observed at any
concurrency level, confirming that the first-writer-wins
Postgres claiming mechanism provides fair load distribution
without a dedicated scheduler.


\subsection{Execution Overhead}\label{sec:eval:overhead}

\name{} generates, compiles, and executes simulation programs
automatically from declarative configuration JSON. This pipeline
introduces overhead relative to a pre-compiled MGPUSim binary
because each simulation requires code generation and compilation
in addition to simulation execution. We characterize this overhead
and show that it is dominated by \texttt{go build} compilation
time, a cost shared with any MGPUSim workflow that modifies a
hardware configuration, and becomes proportionally smaller as
simulation time grows, falling below 10\% for realistic simulation
workloads.

We measure wall-clock time for four benchmarks (MM, NW, ReLU, SC)
across four CU configurations (1, 8, 64, 128 CUs), each run 3
times, for 144 total measurements. These benchmarks span the full
range of simulation durations: MM represents short compute-bound
runs, NW represents medium memory-bound runs, and ReLU and SC
represent long memory-bound runs. The four CU counts capture
behavior at both ends of the topology scale. We compare
against a baseline MGPUSim runner that executes a pre-compiled
simulation binary, representing the best-case execution time with
no code generation or compilation overhead. We decompose each
\name{} run into three components: configuration generation (the
time to translate the configuration JSON into \texttt{main.go},
the only cost unique to \name{}), compilation (\texttt{go build},
a cost shared with any MGPUSim workflow that recompiles after a
configuration change), and simulation execution (identical to the
baseline). We report two overhead scenarios: \textbf{warm start},
which uses a cached Go module cache, and \textbf{cold start},
which includes module download and first-time compilation on a
new machine.

\begin{figure}[t]
    \centering
    \includegraphics[width=\columnwidth]{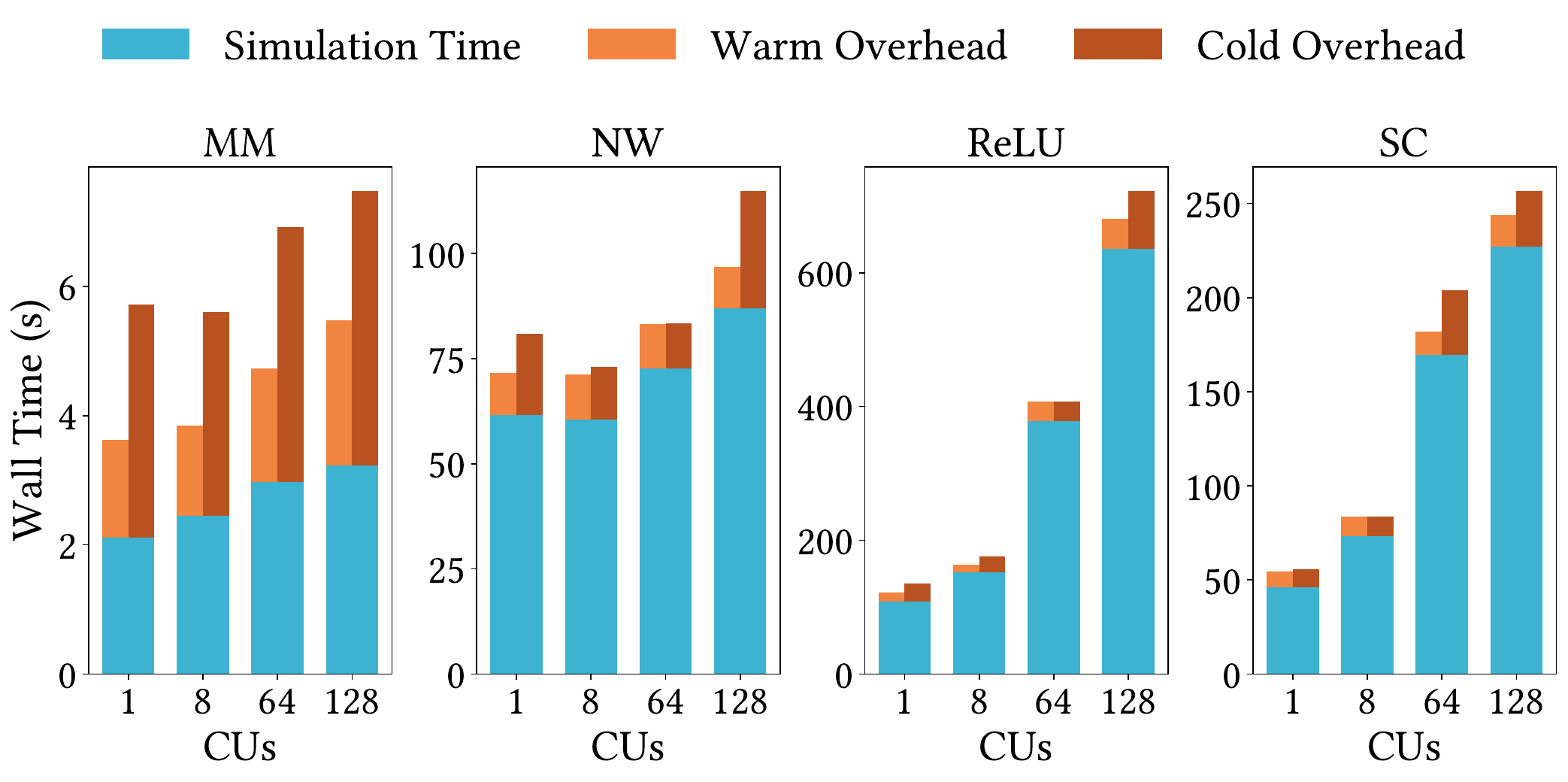}
    \caption{Wall-clock time decomposed into simulation execution
    time, warm start overhead (config generation + cached
    compilation), and cold start overhead (config generation +
    module download + first-time compilation) across four benchmarks
    and four CU configurations.}
    \label{fig:overhead}
\end{figure}

Figure~\ref{fig:overhead} shows wall-clock time decomposed into
simulation time, warm overhead, and cold overhead across all four
benchmarks and CU configurations.

\textbf{The true \name{} overhead is 1.6 seconds per simulation.}
Configuration generation takes a median of 1.6 seconds
(range 0.7--2.3s) across all configurations. All remaining
overhead (warm: median 10.4s additional, cold: median 15.8s
additional) is \texttt{go build} compilation time, which any
MGPUSim workflow pays when modifying and recompiling a
configuration.

\textbf{Overhead decreases with simulation length.}
For short simulations under 10 seconds (MM at 1--8 CUs,
with 2.1--3.2s baseline), overhead reaches 64.3\% because
the 1.6s config generation and 10s compilation dominate a
2--3 second simulation. However, such short simulations are
not representative of architectural studies. For medium
simulations (10--100s, NW and SC at low CU counts), warm
overhead is 15.5\%. For realistic workloads of 100 seconds
or more (ReLU and SC at 64--128 CUs), warm overhead falls
to 8.1\%, with the \name{}-specific configuration generation
cost accounting for just 0.25\% of total wall time. Excluding
MM, warm overhead is 11.8\% mean (11.6\% median) and cold
overhead is 19.1\% mean, with overhead consistently
decreasing as simulation time grows.

The overhead \name{} introduces is the price of eliminating
30 hand-written simulation programs, per-machine environment
setup, orchestration scripts, and custom analysis pipelines
for a $5 \times 12$ study. For studies where simulation time
dominates, which is the common case in architectural research,
\name{}'s overhead is negligible.

%% file: sections/05_Case_studies.tex
\section{Case Study}\label{sec:casestudy}

We demonstrate \name{}'s analysis capabilities through a case study that investigates the performance impact of cache hierarchy configurations across multiple benchmarks and compute-unit (CU) counts. The analysis is performed entirely through \name{}’s interface, without writing any custom parsing or aggregation scripts.

Modern GPU architectures balance compute throughput against memory
system complexity. Adding intermediate cache levels, such as a
shared L1.5 between per-CU L1 caches and a global L2, or an L3
last-level cache below L2, can reduce traffic to main memory but
introduces additional latency on the critical path. Understanding
when deeper hierarchy helps versus hurts requires sweeping multiple
configurations across workloads with different memory behaviors.
Traditionally, each configuration requires modifying simulator
source code and rebuilding; \name{} reduces this to editing a
declarative JSON configuration.

We define four architectural configurations by varying the cache
hierarchy between the per-CU L1 caches and the ideal memory
controller. The baseline routes requests through L1 $\rightarrow$
L2 $\rightarrow$ Memory Controller. The \textbf{+L1.5} variant
inserts a shared 4-way 64B-block L1.5 cache between L1 and L2.
The \textbf{+L3} variant appends a 16-way 128B-block L3 cache
below L2. The \textbf{Combined} variant includes both, forming
L1 $\rightarrow$ L1.5 $\rightarrow$ L2 $\rightarrow$ L3
$\rightarrow$ Memory Controller.

All four configurations use the same compute unit design, TLB
hierarchy, and ideal memory controller. The only difference is
the depth and structure of the data cache path. Each configuration
is expressed as a self-contained \name{} JSON file; no simulator
code changes are required to move between them. Crucially, these
configurations are not parameter variations over a fixed topology:
the \textbf{+L1.5} and \textbf{+L3} variants differ in component
presence (the L1.5 and L3 nodes are absent in the baseline),
associativity (4-way vs.\ 16-way), block size (64B vs.\ 128B),
and structural wiring, each expressed as a declarative graph edit.
This demonstrates that \name{}'s configuration abstraction
supports structural composition from the imported Models, not only
parameter sweeps, across a space that spans component presence,
integer-valued settings, and interconnect topology simultaneously.

We select three benchmarks that stress different parts of the
memory system: ReLU, a streaming element-wise operation that is
primarily memory-bandwidth bound; Matrix Multiply (MM), a
compute-intensive kernel with high arithmetic density relative
to memory traffic; and Matrix Transpose (MT), a stride-heavy
access pattern that stresses cache line utilization and memory
bandwidth. We run each configuration at three CU counts (1, 8,
and 16) and at two problem-size tiers: a smaller tier suitable
for local iteration and a larger tier representative of realistic
workloads. The full experiment comprises 72 simulation runs, each
launched from the same runner binary without modifying any
simulator source code.

The current case study targets the MGPUSim/Akita ecosystem and
three benchmark families. This reflects the scope of the current
code generation backend rather than a limitation of the
underlying experiment abstraction: the core concepts of
configurations, experiment matrices, runners, and the artifact
store are defined independently of any specific simulator, and
extending the evaluation to additional simulators such as gem5
or SST is a priority for future work. Within the MGPUSim backend, the configuration interface additionally supports sweeping interconnect bandwidth, TLB size,
memory controller count, and CU pipeline parameters, though a
comprehensive evaluation across all these dimensions is left to
future work.


\begin{figure}[t]
    \centering
    \begin{subfigure}[t]{\linewidth}
        \centering
        \includegraphics[width=\linewidth]{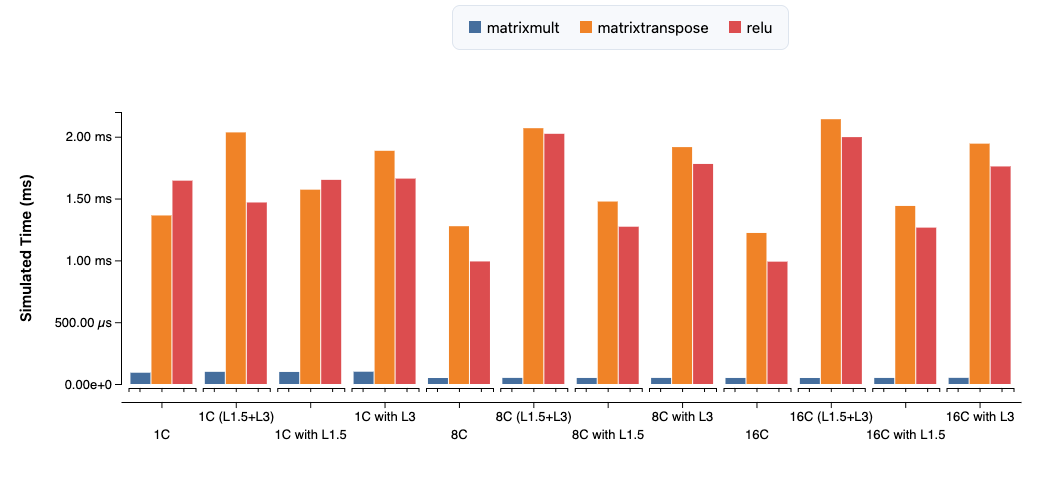}
        \caption{Experiment-level view for the cache hierarchy case study. Each group of bars corresponds to a (configuration, CU count) combination; bar height represents simulated execution time. The three benchmark series (matrixmult, matrixtranspose, relu) are shown in separate colors. The view enables direct identification of configurations where deeper cache hierarchies increase rather than decrease execution time.}
        \label{fig:case_overview}
    \end{subfigure}

    \vspace{0.4em}

    \begin{subfigure}[t]{\linewidth}
        \centering
        \includegraphics[width=\linewidth]{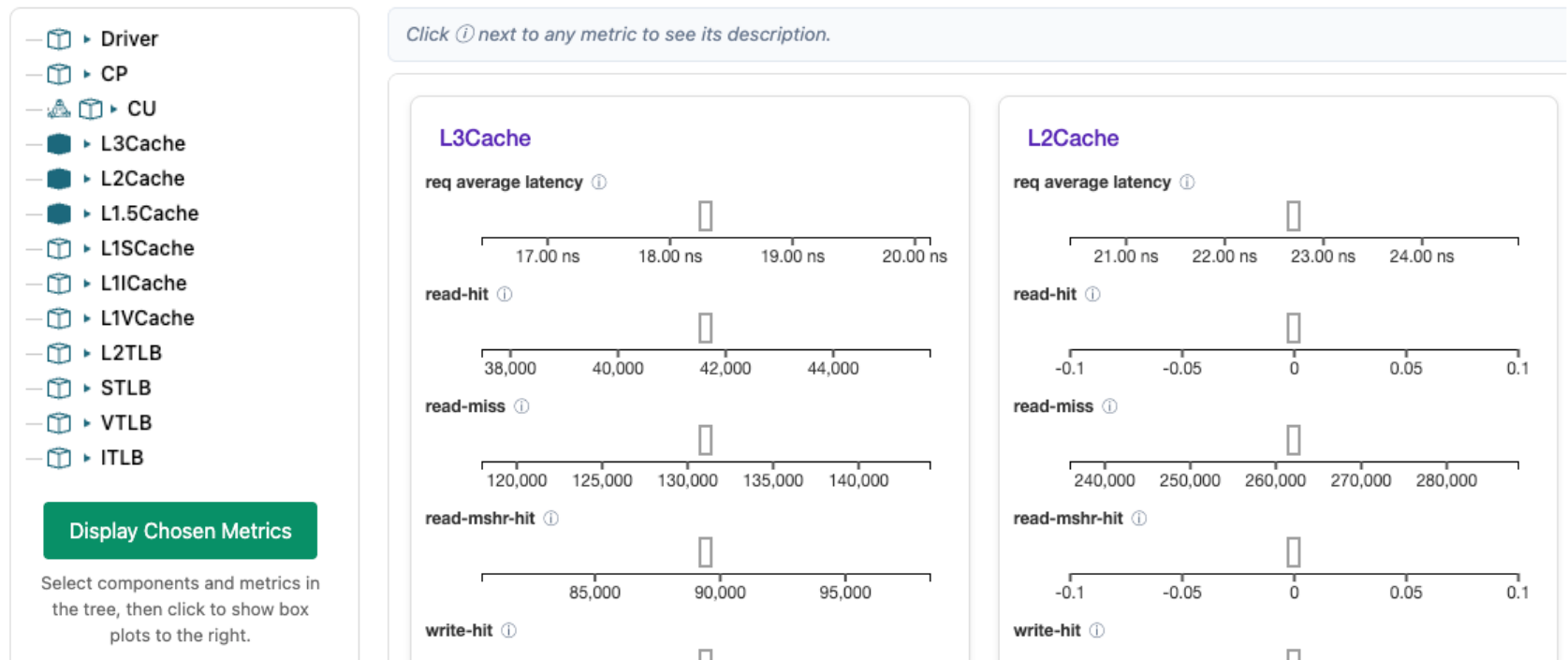}
        \caption{Component-level drill-down for the matrixtranspose benchmark on the 1C (L1.5+L3) configuration. The left panel shows the component hierarchy; the right panels show per-component metric timelines. The near-zero L2 read hit rates and nonzero L3 MSHR activity visible here explain the performance degradation identified in the experiment-level view. 
        }
        \label{fig:case_drilldown}
    \end{subfigure}
    \caption{Case-study analysis views in ArchSim.}
    \label{fig:case_analysis}
\end{figure}

Once the experiment completes, we navigate to the Analysis tab
to explore the results. At the top of the tab, the
experiment-level view (Figure~\ref{fig:case_overview}) presents
simulated execution time across all
$(\textit{configuration}, \textit{benchmark})$ pairs, enabling
direct comparison of performance trends across the entire
experiment matrix.


Several insights can be immediately observed.
\begin{itemize}[nosep,leftmargin=*]
    \item \textbf{Workload sensitivity to memory hierarchy.}
\texttt{matrixmult} exhibits consistently low execution time
across all configurations. This is consistent with its
compute-intensive nature and relatively small problem size
(128$\times$128$\times$128), which limits its sensitivity to
cache hierarchy changes. In contrast, \texttt{matrixtranspose}
and \texttt{relu} show substantial variation across
configurations, suggesting stronger dependence on memory-system
behavior due to their stride-heavy and streaming access patterns
respectively. 
    \item \textbf{Deeper cache hierarchies do not universally improve performance.}
Configurations that include both L1.5 and L3 caches frequently exhibit higher execution time than simpler configurations. This indicates that additional cache levels may introduce overhead that outweighs their benefit for certain workloads.
    \item \textbf{Interaction between parallelism and memory hierarchy.}
Increasing CU count does not uniformly improve performance
across all configurations. For configurations with deeper cache
hierarchies such as \textbf{Combined} (L1.5+L3), performance
at higher CU counts is sometimes higher than at lower CU counts,
suggesting that the latency overhead of deeper hierarchies can
outweigh the benefits of increased parallelism for certain
workloads.
\end{itemize}
This view allows researchers to quickly identify anomalous or non-intuitive performance trends without constructing experiment-specific aggregation pipelines.

To explain the observed performance differences, we select a
representative configuration: \textit{matrixtranspose on 1C
(L1.5+L3)}. Selecting this point in the experiment-level view
loads the component-level drill-down interface, which exposes
hierarchical metrics for each hardware component.
Figure~\ref{fig:case_drilldown} shows an excerpt of this view,
focusing on the L2 and L3 cache metrics most relevant to the
observed performance degradation.

The drill-down reveals the following: 

\begin{itemize}[nosep,leftmargin=*]
    \item \textbf{Ineffective intermediate cache level.}
The L2 cache exhibits near-zero read and write hit rates, indicating that it provides little filtering benefit for this workload. Most memory accesses pass through this level without reuse. 
    \item \textbf{Useful deeper cache level.} The L3 cache shows nonzero read-hit and MSHR-hit activity, indicating that some reuse is captured at this level. 
    \item \textbf{Explaining performance degradation.}
    The L2 cache adds 21--24\,ns of average request latency
while providing effectively zero hits, so every request pays
this latency cost before reaching the L3. The combined
overhead of traversing both levels without proportional
reuse benefit explains the increased execution time observed
in the experiment-level view.
\end{itemize}


This analysis demonstrates how \name{} enables direct causal reasoning: from identifying a performance anomaly at the experiment level to diagnosing its architectural root cause at the component level.

%% file: sections/07_Related_work.tex
\section{Related Work}

Widely used frameworks such as gem5~\cite{binkert2011gem5},
Multi2Sim~\cite{ubal2012multi2sim,ubal2007multi2sim,gong2017multi2sim},
GPGPU-Sim, Sniper~\cite{carlson2011sniper},
SST~\cite{rodrigues2011structural}, and
MGPUSim~\cite{sun2019mgpusim} provide detailed models for
evaluating CPU and GPU architectures. These systems focus on
modeling fidelity and execution performance; experiment structure
remains implicit in the scripts and file organization that
surround them. \name{} builds on this ecosystem by introducing
an experiment platform that represents simulation studies as
structured entities and automatically generates the simulator
programs required to execute them.



ReproZip~\cite{chirigati2016reprozip} captures execution
environments to support experiment replication. OCCAM~\cite{oliveira2018occam} provides a software environment for creating reproducible research artifacts.  MLflow~\cite{zaharia2018accelerating}
and Sacred~\cite{greff2017sacred} provide experiment tracking for
machine learning workflows. Nextflow~\cite{di2017nextflow},
Snakemake~\cite{koster2012snakemake}, and
Kepler~\cite{altintas2004kepler} support reproducible pipelines
in data-intensive science. These systems target flat parameter
spaces and scalar outputs and do not represent architecture
simulation structure: hardware topology graphs, typed port
bindings, and spatially organized component-level metrics.
Within the architecture community, gem5art~\cite{bruce2021enabling}
captures configuration scripts, disk images, and simulator binary
versions using Git-based artifact tracking. gem5art is the
closest prior work to \name{}'s version preservation design, but
requires the researcher to explicitly tag each run with a version
identifier and does not propagate version information to
subsequent runs in the same experiment. \name{} records simulator
dependency versions automatically on the first successful run and
reuses them for all subsequent runs without researcher
intervention.

Daisen~\cite{sun2021daisen}, Vis4Mesh~\cite{wang2023visual},
Ziabari et al.~\cite{ziabari2015visualization}, and Ariel
et al.~\cite{ariel2010visualizing} provide fine-grained visualization
of GPU and accelerator execution behavior using simulation traces. 
gem5 also provides hierarchical statistics
collection~\cite{binkert2011gem5}. These tools analyze individual
simulation runs; cross-run analysis requires custom parsing scripts.
\name{} operates at the experiment level, storing outputs as structured artifacts that support experiment-scoped metric retrieval and component-level exploration without intermediate parsing pipelines.

%% file: sections/08_Conclusion.tex
\section{Conclusion}

Computer architecture research increasingly relies on large-scale
simulation studies that explore complex design spaces across many
configurations and workloads. Yet the infrastructure supporting
these studies evolves through collections of scripts, configuration
programs, and analysis pipelines whose relationships remain implicit
in code and directory structures, making studies fragile, difficult
to reproduce, and hard to scale. \name{} addresses this by treating
simulation experiments as structured infrastructure objects: by
making experiment structure explicit and machine-readable, the
platform automates execution, artifact preservation, and analysis
across the experiment lifecycle while preserving full compatibility
with existing architecture simulators.

Looking forward, we envision simulation experiments becoming
portable and shareable artifacts rather than collections of scripts
tied to a particular environment. Such a shift would enable
researchers to reuse architectural studies, reproduce prior results
with minimal manual effort, and build shared infrastructure for
exploring emerging architectural ideas. \name{} represents one
step toward this direction, and we believe that elevating experiment
structure to a first-class concept can fundamentally improve how
architecture simulation studies are created, executed, and shared.